# A Topological Encoding Method for Data-driven Photonics Inverse Design

*Zhaocheng Liu*[1,2*], *Zhaoming Zhu*[1], *Wenshan Cai*[2]

[1] Facebook, Inc.

[2] School of Electrical and Computer Engineering, Georgia Institute of Technology, Atlanta, Georgia 30332
zcliu@gatech.edu

**Abstract:**

Data-driven approaches have been proposed as effective strategies for the inverse design and optimization of photonic structures in recent years. In order to assist data-driven methods for the design of topology of photonic devices, we propose a topological encoding method that transforms photonic structures represented by binary images to a continuous sparse representation. This sparse representation can be utilized for dimensionality reduction and dataset generation, enabling effective analysis and optimization of photonic topologies with data-driven approaches. As a proof of principle, we leverage our encoding method for the design of two dimensional non-paraxial diffractive optical elements with various diffraction intensity distributions. We proved that our encoding method is able to assist machine-learning-based inverse design approach for accurate and global optimization.

## 1. Introduction

Over the past two decades, the advancement of photonics has been enabling vast approaches for manipulating light in the wavelength scale. By engineering the building blocks of the photonics materials and devices, the behavior of light, such as phase, amplitude and polarization of transmitted light and near field responses of wave, can be accurately controlled. Diverse photonic devices ranging from diffractive optical elements (DOE) to metamaterials (MM) and metasurfaces (MS)[1,2], are designed for the extensive applications such as virtual/augmented reality display[3,4], miniaturized imaging systems[5,6], and quantum optics platform[7]. However, the complex mechanism of light-matter interaction prevents an intuitive strategy for the design of the building blocks in the photonic devices. As such, various inverse design and optimization algorithms have been developed for the expeditious design of photonic structures. For example, given some parameters that define the photonic device, adjoint methods[8,9] calculates the gradients of the parameter with respect to the objective function, and incrementally updates the structure by subtracting the gradients from the parameters. Metaheuristic optimizations[10,11], on the other hand, treat the physical system as a black box, and update the parameters following the manner inspired by physical and biological systems.

With the fast evolution of machine learning (ML) and deep learning (DL) techniques[12], data-driven methods are emerging as an alternative way to discover and design photonic structures and devices[13]. Feedforward neural networks are leveraged for the approximation of the photonics systems with tens to hundreds of parameters, and have been utilized to successfully optimize photonic structures such as photonic crystals[14,15], waveguide[16,17], chiral metamaterials[18], and metasurfaces[19]. When the degree of freedom (DOF) of the photonic system grows to thousands and more, convolutional neural networks (CNN) are adopted for the accurate prediction of the physical responses with much lower computational complexity[20,21]. Photonic structures represented in pixelated images, for example, are usually processed by CNNs to reduce the DOF for further optimization. Additionally, generative models, such as variational autoencoder (VAE)[22] and generative adversarial network (GAN)[23,24], are utilized for the design of high DOF metasurface nanostructures in an expeditious way[25-28]. The stochastic nature of the generative models enables the exploration of the solution space in a global way. Consolidating with traditional optimization techniques, GAN and VAE are able to discover the topology of nanostructures with improved efficiency and robustness[29-31].

In the problems of inverse design of photonic structures, optimizing the topology of a photonic structure with arbitrary shape is a long-sought-after goal. Typically, the topology of photonic structures is represented in binary images. Because of the discretization and the high DOF of binary images, optimization is likely stuck in local minima. Although generative models are able to discover new topologies so as to approach global minimum, the bias of the training dataset and the limited capacity of the network cause incompleteness of the solution space, i.e., the global minimum may not be included in the space defined by the training dataset. Here, we propose an encoding method that is able to transform the binary image to a continuous sparse representation. This encoding approach can be used for data generation and dimensionality reduction of photonic structures, augmenting the capacity of ML models for analyzing the dataset, and enhancing the likelihood of achieving the global minimum in the optimization problems. As a proof of principle, we consolidate the proposed encoding method and a DL-based optimization framework[26,32,33] to inverse design and optimize non-paraxial diffractive optical elements (DOE) with various diffraction intensity distributions. Traditionally, iterative Fourier transform algorithm (IFTA) is used for the design of binary phase mask of the DOE with small diffraction angle[34,35]. However, this approach does not take into account the physical process and loses its fidelity when the paraxial approximation is not valid. We proved that our encoding method is able to assist data-driven methods for the accurate and global optimization of DOE topologies.

**2. Encoding the topology of photonic structures**
Our goal is to encode the nanostructures represented in a 2D binary image to a continuous sparse representation so as to assist data-driven approaches such as ML and DL to analyze, discover, and optimize the topology of photonic structures. In the following discussion, we denote the binary image to be encoded as $f(x,y) \in \{0,1\}^N$, where $x$ and $y$ are coordinates of the image, and $N$ is the dimension of the image. Fourier transform (FT) can be used to transform the binary image to a sparse representation. However, when some operations such as filtering out high-frequency components, are applied to the transformed sparse representation, the inverse Fourier transform (IFT) of the sparse representation may not be a binary image anymore. Consequently, FT is not able to encode, decode, and manipulate arbitrary topologies of photonic structures for the general purpose of dimensionality reduction, data analysis, and device optimization.

Instead of simply applying FT to the binary image, we carry out FT to the level set function $\phi(x, y)$ that defines the topology of the structure. As illustrated in Fig. 1(a), a level set function $\phi(x, y)$ is defined as a 3D surface, and the topology of the structure (encircled by red line) can be represented as the zero-level set of $\phi$ as $\Gamma = \{(x, y)|\phi(x, y) = 0\}$. Given certain binary images as shown in Fig. 1(b), our encoding strategy is first to construct a level set function, and then apply FT to the level set function so as to derive the sparse representation of the binary image as shown in Fig. 1(c). Reversing the whole process reconstructs a sparse representation to a binary image. The detailed procedure of encoding and decoding photonic structures are shown in Fig. 1 (d). We first construct the level set function of the original image $f(x, y)$ through the transform:

$$\phi_e(x, y) = e^{i\pi f(x,y)} \tag{1}$$

The exponential function maps the original image to the discrete value $\{-1,1\}^N$, so that the topology of the structure can be represented by $\Gamma = \{(x, y)|\phi_e(x, y) = 0\}$. By carrying out FT to $\phi_e$, we can find the sparse representation of $\phi_e$ in the complex frequency space as

$$\hat{\phi}_e(k_x, k_y) = \mathcal{F}[\phi_e(x, y)] \tag{2}$$

By the property of FT, $\hat{\phi}_e$ naturally satisfies the condition:

$$\hat{\phi}_e(k_x, k_y) = \hat{\phi}_e^\dagger(-k_x, -k_y) \tag{3}$$

where $\hat{\phi}_e^\dagger$ is the conjugate of $\hat{\phi}_e$. Thus, for an image with $N$ pixels, the DOF of its sparse representation is also $N$. Certain operations $P$ such as filtering out high-frequency components can be performed on this sparse representation $\phi_d(x, y) = P\phi_e(x, y)$. In order that the decoded image from $\phi_d(x, y)$ is also binary, $\phi_d$ should meet the same condition $\hat{\phi}_d(k_x, k_y) = \hat{\phi}_d^\dagger(-k_x, -k_y)$. To decode the image, we apply the encoding process in a reversed order. In detail, we first apply inverse Fourier transform (IFT) to $\hat{\phi}_d$ to find the level set function that represents the photonic structure

$$\phi_d(x, y) = \mathcal{F}^{-1}[\hat{\phi}_d(k_x, k_y)] \tag{4}$$

and then retrieve the binary image representation through the operation:

$$f_d(x, y) = \frac{1}{\pi} \text{ang}(\phi_d(x, y)) \tag{5}$$

where ang(·) is a function that calculates the phase of $\phi_d(x, y)$, and $\frac{1}{\pi}$ is the normalizer that ensures the retrieved image has binary values 0 and 1. Note Eq. (4) is identical to setting a threshold $\phi_0 = 0$ and retrieving the image by

$$f_d(x,y) = \begin{cases} 1 & if \ \phi_d(x,y) > \phi_0 \\ 0 & otherwise \end{cases} \tag{6}$$

## 3. Properties of the encoding method

The encoding method illustrated above transforms a binary image to a continuous sparse representation, allowing incremental variation of the topology of the structures by perturbing $\hat{\phi}_d(k_x, k_y)$. We will illustrate several properties of this encoding method and show its advantages in representing photonic structures for data-driven photonic discovery. To be consistent with terminology in machine learning, we will call the space of the sparse representation as latent space and $\hat{\phi}_d$ as latent vector.

### a. Dimensionality reduction

Suppose a photonic structure represented in an image with number of pixels $N$, the DOF of the structure is also $N$. Performing inverse design and optimization in such a high-dimensional space is difficult. On the other hand, the topology of a structure usually presents some properties such as continuity and connectiveness for the purpose of proper simulation and fabrication. Thus, the available structures cluster in a small region in the $N$-dimensional image space. Machine learning algorithms such as VAE and GAN have been used to reduce the dimensionality of the photonic structures. However, a trained ML model can only faithfully encode and decode the structures topologically similar to the training dataset. Our method, derived without the dependence of data, is able to process binary image data in a fast and general manner.

Figure 2(a) to (d) illustrate the process of encoding and decoding a photonic structure. The initial structure, as shown in Fig. 2(a), is represented in an image with $N = 64 \times 64$. Our method encodes the structure into its frequency representation $\hat{\phi}_e$, the norm of which is presented in Fig. 2(b). After cropping the high-frequency components as in Fig. 2(c), only dominant components are kept in $\hat{\phi}_d$. Recovering $\hat{\phi}_d$ into binary images through our method, we achieve the decoded image. Interestingly, as decoding the structure is the essentially performing the IFT, the recovered images can have arbitrarily large DOF. The decoded example shown in Fig. (d) has a resolution of $128 \times 128$. As cropping inevitably deletes portion of the information in the latent space, our encoding method is an irreversible lossy compression. The higher order of terms kept in the latent space, the finer feature will be remained in the decoded images. Other dimensionality reduction approaches can be applied to the encoded latent vectors to further reduce the complexity of the inverse design problem.

### b. Continuity of the latent space

Since FT and IFT are uniformly continuous operators, the decoded image $f_d(x,y)$ can be incrementally varied by perturbing $\hat{\phi}_d(k_x, k_y)$. Figure 2(e) to (i) shows a continuous topological variation from the first pattern shown in Fig. 2(b) to the last one in Fig. 2(f). The first and last patterns are randomly constructed with latent vectors $\hat{\phi}_1, \hat{\phi}_2 \in \mathbb{R}^{7 \times 7}$. Figure 2(c) to 2(e) present the intermediate that are decoded from the linearly interpolated latent vector $\hat{\phi}_d = \lambda\hat{\phi}_1 + (1-\lambda)\hat{\phi}_2$, where $\lambda \in (0,1)$. As we can observe, the two distinct topologies can be smoothly transformed by linearly interpolating their latent vectors. This property is indispensable for the fast convergence when evolution strategy (ES) is utilized for the topology optimization [32]. It is noteworthy that linear interpolation does not always result in continuous topological transformation, especially when the latent space is high dimensional and the topologies of the two patterns are significantly distinct. In this situation, geodesic, representing the shortest path between the two patterns in the latent space, should be computed for the smooth transformation.

### c. Symmetric property

Symmetry is a crucial geometric property that should always be considered in the design of photonic devices. Properly leveraging the symmetric property reduces the time of simulation and mitigates the difficulty of optimization. Our encoding method maintains the symmetric properties of binary images in the frequency spaces. Figure 2(j) to (n) display a few randomly generated patterns with some symmetric properties. The dimension of the latent space we chose is $N = 5 \times 5$. Without any constraints on the latent space, the generated pattern shown in Fig. 2(j) does not present any symmetry. In order to generate a centrosymmetric pattern, we need to enforce the latent vector to be centrosymmetric, i.e., $\hat{\phi}_d(k_x, k_y) = \hat{\phi}_d(-k_x, -k_y)$. Combined with Eq. (3), this condition is equivalent to the latent vectors being real, reducing the DOF of the latent vectors to $\lfloor N/2 \rfloor + 1 = 13$. Figure 2(k) is a randomly generated centrosymmetric pattern with such constrain. Similarly, if $\hat{\phi}_d(k_x, k_y) = \hat{\phi}_d(k_x, -k_y)$ and $\hat{\phi}_d(k_x, k_y) = \hat{\phi}_d(-k_x, k_y)$ are enforced, we can generate axisymmetric patterns such as the one shown in Fig. 2(l). The DOF in this case is reduced to $\left(\lfloor \sqrt{N}/2 \rfloor + 1\right)^2 = 9$. By additionally constraining $\hat{\phi}_d(k_x, k_y) = \hat{\phi}_d(k_y, k_x)$, axisymmetric patterns with axis of symmetric $y = x$ can be produced as shown in Fig. 2(m) and (n). In this circumstance, the DOF of the pattern is $\left(\lfloor \frac{\sqrt{N}}{2} \rfloor + 1\right)\left(\lfloor \frac{\sqrt{N}}{2} \rfloor + 2\right)/2 = 6$, indicating that only six variables are required for arbitrarily manipulate the topology of photonic structure. Such symmetric properties of the encoding/decoding method can be leveraged for reducing the parameters in the inverse design of metasurfaces and photonic crystals with specific polarization requirements.

### d. Multilevel optimization

When the DOF of a photonic structure is large, optimization techniques suffer from problems such as slow convergence and local minimum. In this situation, multilevel optimization[36] can be used for designing the structure and enhancing the performance. By the nature of FT, our encoding method allows the multilevel optimization of photonic structures by gradually modifying the corresponding latent vectors. Figure 2(o) to 2(s) present an example that adding finer features to the initial structure (Fig. 2(o)) through our encoding method. The initial structure is constructed from a $\hat{\phi}_d^0 \in \mathbb{R}^{7 \times 7}$. By attaching additional vectors to the latent vector, we can augment the latent vector to $\hat{\phi}_d^1 \in \mathbb{R}^{9\times 9}$ in a higher dimension. Figure 2(p) presents the decoded image of the augmented latent vector $\hat{\phi}_d^1$. A few features such as the hole in the center appear. Repeating the augmentation process results in the incremental evolution of the structures with finer features as shown in Fig. 2(q) to 2(s). This unique property of the encoding method enables the consolidation of traditional optimization and multilevel optimization for the inverse design of high DOF photonic structures.

## 4. Designing non-paraxial diffractive optical elements (DOEs)

As a case study, in this section we will represent how the encoding method can be applied in the inverse design and optimization of binary DOEs. Traditionally, the design of DOEs are relied on iterative Fourier transform algorithm (IFTA). The algorithm can generate binary phase masks whose FT is proportional to the required diffraction intensity distributions. However, IFTA does not take into account the actual physical process, resulting in the inaccuracy of the modeling for diffraction intensity. This inaccuracy prevents an effective design method for the few dots, non-paraxial diffractive beam splitters. To solve this problem, we consolidate the proposed encoding method and a hybrid inverse design algorithm[32] to design $3 \times 3$ non-paraxial diffractive beam splitters with various diffraction intensity distributions.

The cross section of common configuration of the DOEs is shown in Fig. 3(a). The grating and the substrate share the same materials with a refractive index $n = 1.56$, and the thickness of the grating pattern is $t = 840\ nm$. Monochromatic light with a wavelength of $\lambda_0 = 940\ nm$ is incident from the substrate side. We set the period of the DOE as $p = 2.83\ \mu m$, so that the angle between the 1st and the 0th order diffraction is about 21°. The objective of the design is to identify the DOE patterns that are able to accurately diffract the incident light into the 9 different directions with various required intensity distributions.

In order to leverage data-driven approaches such as DL for the fast and global optimization, we first generate sufficient DOE that are able to diffract light to the desired directions. Since the encoding method we proposed is based on FT, it is sufficient to sample $\hat{\phi}_d$ from $[-1,1]^{3\times 3}$ as the sparse representation for the design of $3 \times 3$ diffractive beam splitters. In practice, we can further simplify the representation of each DOE for the convenience of the training of a neural network model. In detail, we write $\hat{\phi}_d$ as:

$$\hat{\phi}_d = \begin{bmatrix} v_1 + iv_2 & v_3 + iv_4 & v_5 + iv_6 \\ v_7 + iv_8 & v_9 & v_7 - iv_8 \\ v_5 - iv_6 & v_3 - iv_4 & v_1 - iv_2 \end{bmatrix} \quad (7)$$

where $\{v_i | i = \{1, \ldots, 9\}\}$ are real numbers. Equation (7) satisfies the condition defined by Eq. (3) and has the DOF of 9. Sampling a random $\hat{\phi}_d$ is achieved by independently sampling each $v_i$ from a uniform distribution. We reorganize the entries in $\hat{\phi}_d$ to an encoded vector $v = [v_1, v_2, \ldots, v_9]$, and take the encoded vector as the input of the network. For each encoded vector $v$, we simulate the diffraction efficiencies $\eta$ of corresponding grating structures with rigorous coupled wave analysis (RCWA)[37]. Under the wavelength of incident light $\lambda_0 = 940\ nm$, the DOEs produce

5 × 5 orders of diffraction light. Next, the simulated diffraction intensities are reorganized into a 25-dimensional vector, and normalized with respect to the largest intensity $\eta_{max}$. The final representation of each simulated result is denoted as a 26-dimensional vector:

$$K = \left[\frac{\eta_{-5,-5}}{\eta_{max}}, \frac{\eta_{-5,-4}}{\eta_{max}}, \dots, \frac{\eta_{5,5}}{\eta_{max}}, \eta_{max}\right] \quad (8)$$

As such, (*v*, *K*) is a training pair of the network model. To construct the whole dataset for the training of the network, we randomly sampled 15,000 encoded vectors *v* and performed the process outlined above. We split the dataset into two parts with 12,000 for training and the rest for validation. Since the grating patterns are represented in low-dimensional vectors *v*, simple neural network architecture is sufficient for accurate approximation of the diffraction efficiency. As shown in Fig. 3(b), we build an eight-layered fully connected neural network with input of encoded vector *v* and output of efficiency vector *K*. All the hidden layers of the network contain 128 neurons, and the nonlinear activations after each input and hidden layers are ReLU. Figure 4(a) presents the loss variation during the training process. The validation loss below 0.03 is achieved after 100 epochs of training.

In order to globally optimize the topology of the grating patterns, we adopt the modified evolution strategy (ES) [32] as shown in Fig. 3(c). The algorithm starts with sampling a population of random encoded vector *v*. Each vector is regarded as an individual in the population. These vectors are simulated through the neural network simulator. Based on the simulated results, the population is subsequently evaluated by certain design objective, and the elites of the population are selected for the following reproduction and mutation. The algorithm iterates until one of the individuals achieve the design criterion or the maximum iteration is reached. To design DOEs with various diffraction intensity distributions $\eta_{obj} \in [0,1]^{3\times3}$, we define the uniformity error of a design as:

$$U_{err} = \frac{\tilde{\eta}_{max} - \tilde{\eta}_{min}}{\tilde{\eta}_{max} + \tilde{\eta}_{min}} \quad (9)$$

where $\tilde{\eta}_{max}$ and $\tilde{\eta}_{min}$ are the maximum and minimum intensity of the scaled diffraction intensity $\eta_{scaled} = \frac{\eta}{\eta_{obj}}$, where $\eta$ is the simulated performance of the designed beam splitter. Our objective is to minimize $U_{err}$ of a DOE design given certain intensity distribution $\eta_{obj}$. As neural network is an approximation model, deviation of identified solution is unavoidable. Instead of augmenting the dataset for an improved network simulator, we choose to carry out the ES-based optimization for 150 times, simulate all the identified structures, and select the optimal solution with minimal $U_{err}$. With the acceleration of the network simulator, each run of the ES-based optimization is within 1 second. Figure 4(b) shows the histogram of $U_{err}$ after 150 runs of the optimization when the intensities of all diffraction orders are required to be equal (i.e. all entries in $\eta_{obj}$ are 1). The blue and orange bars represent the distribution of $U_{err}$ evaluated by the network simulator and RCWA respectively.

Figure 5(a) to (h) present eight examples of designed DOE with various intensity distribution $\eta_{obj}$. For each panel in Fig. 5, the left plot is the tiled unit cell of the designed DOE, the middle image shows the simulated diffraction intensities of all orders, and the right plot compares the desired diffraction intensities (blue) and RCWA simulated results of the designed DOE (orange). With the help of our encoding method, the hybrid framework successfully identified DOE structures with diffraction intensity distribution matching the design objectives. Quantitively, the uniformity errors $U_{err}$ of designed DOEs from Fig. 5(a) to 5(h) are 0.035, 0.045, 0.073, 0.068, 0.194, 0.036, 0.352, and 0.079, respectively. Since $U_{err}$ is calculated through scaled intensity $\eta_{scale}$ and objective intensity $\eta_{obj}$ is the denominator of $\eta_{scale}$, the error is extremely sensitive to the objective intensities with small values. In the examples shown in Fig. 5(f), 5(h) and (i), diffraction intensities in some orders are required to be small. In such cases tiny disagreement of the actual diffraction and objective intensities induces large $U_{err}$. Nevertheless, the overall intensity distributions of all designs have excellent agreement with respect to the objectives, confirming the effectiveness of our encoding methods in the machine-learning-based inverse design approaches.

## 5. Conclusion

In summary, we have proposed an encoding method that is able to transform the topology of a photonic structure represented in discrete, high-dimensional, and binary pixelated images into a continuous sparse representation. We explored the properties of this encoding methods, such as continuity of the latent space and symmetric properties, and discussed the potential application of this method in the dimensionality reduction and data generation for the data-driven photonics optimization. As a case study, we utilized the encoding method and a deep learning-based optimization framework to design 3 × 3 DOEs with non-paraxial diffraction angle and various diffraction intensity distributions. The encoding method allows us to generate sufficient data for the optimization without explore unnecessary solution space. The encoded DOE represented in the low dimension also enhances the accuracy of the network and, as a result, increases the fidelity of the design.

Although the proposed encoding methods are aimed to assist data-driven inverse design of photonic structures, other derivative-free traditional optimizations[38] can also take advantage of the continuous low-dimensional representation of photonic structures. If global optimization is not required, local search algorithms can be applied to the latent vectors of the structures without generating redundant dataset. In the future, we expect to explore the application of the encoding method with both traditional and data-driven optimization approaches for the discovery and design of other photonic media such as photonic crystals and metasurfaces, and anticipate to consolidate the encoding method with deep generative models to produce complex patterns for the general inverse design of photonic structures.

# Figures

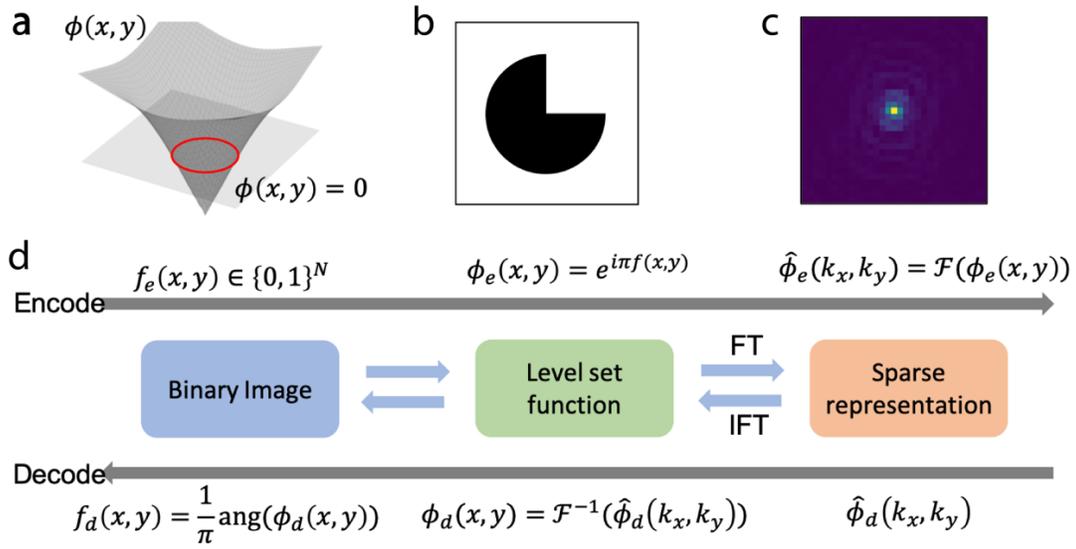

Fig. 1 **Description of the encoding method** | (a) Illustration of a level set function $\phi(x,y)$. The topology (encircled by red line) is represented by the zero-level set. (b) Binary image representation of a nanostructure. (c) Encoded sparse representation of the nanostructure shown in (b). (d) The outline of the encoding method. For the encoding process, a level set function is first constructed from the given binary image. The spare representation of the image is derived from the Fourier transform of the level set function. Decoding the binary image from the sparse representation is the reverse of the encoding process.

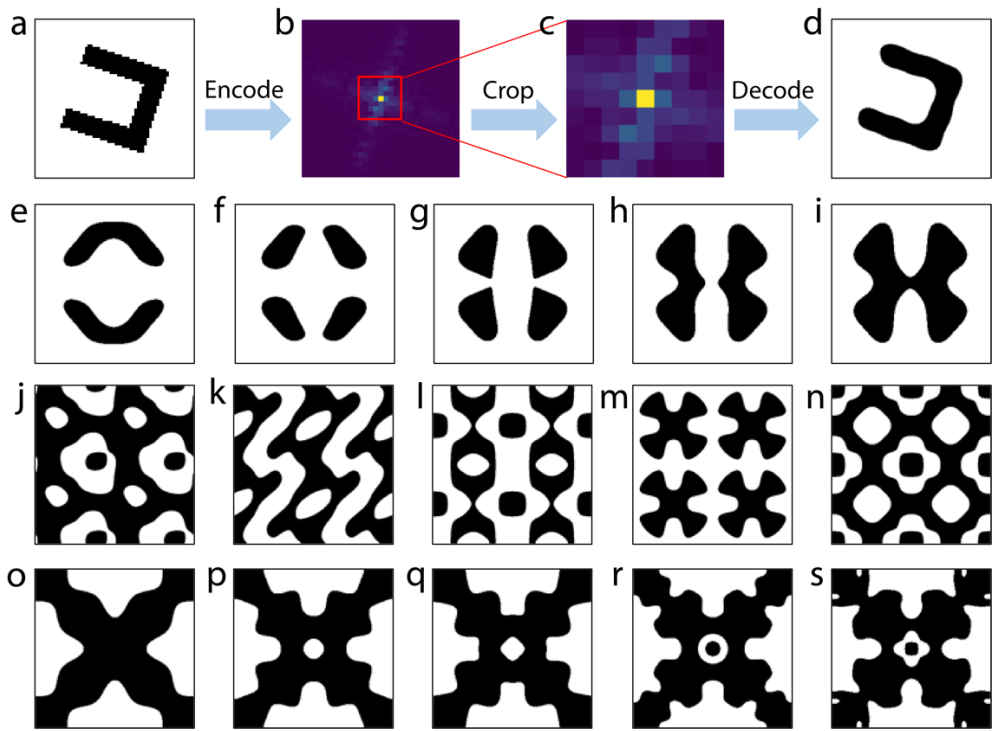

Fig. 2 **Properties of the encoding method** | (a) – (d) Dimensionality reduction using the proposed encoding method. Initial binary image (a) is encoded to the sparse representation (b). The low-dimensional representation (latent vector) can be achieved by deleting the high-frequency components as shown in (c). The latent vector can be recovered to the initial structure without substantial loss of information. (e) – (i) Continuously varying two topologies by linearly interpolating the latent vectors. (j) – (n) Generated samples with various symmetric properties. The shown images are tiled unit cells of the generated patterns. (o) – (p) Adding fine features to initial pattern (o) by gradually expanding the dimensions of latent vectors from $7 \times 7$ to $15 \times 15$.

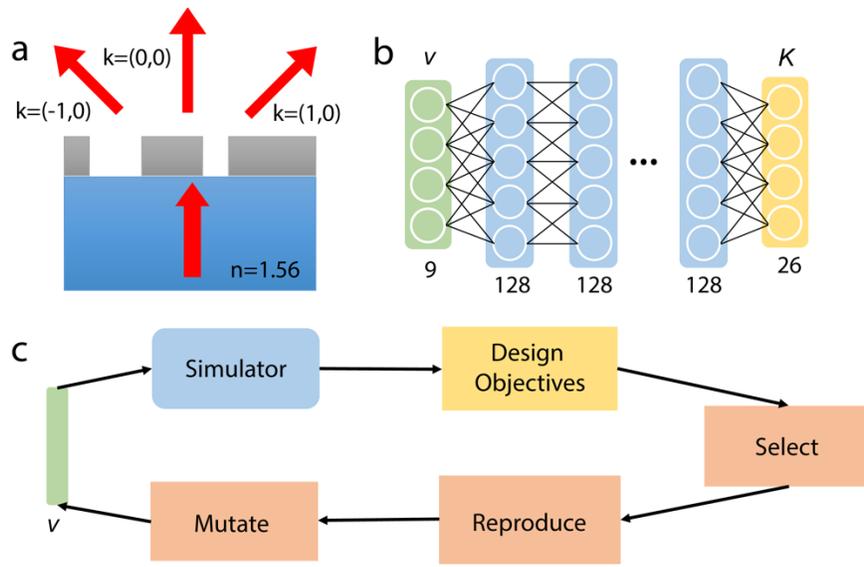

Fig. 3 **Configuration of the DOE and the optimization method** | (a) The cross section of the DOE. The grating pattern and the substrate share the same material with a refractive index of 1.566. The period of the DOE is $p = 2.83\ \mu m$ and the thickness of the grating pattern is $t = 840\ nm$. Light with a wavelength $\lambda_0 = 940\ nm$ is incident from the substrate side. Our aim is to optimize the grating pattern such that the central $3 \times 3$ order diffraction present various intensity distributions. The angle between 0 and +1 order diffraction is 21°. (b) Architecture of neural network simulator for predicting the diffraction intensities of DOEs. The input is the encoded vectors of the DOEs, and the output is the vector containing normalized diffraction intensities and maximum intensity of all diffraction orders. The network is an eight-layer fully connected networks, and each hidden layer has 128 neurons. (c) Schematic of the evolution strategy. Randomly generated latent vectors are evaluated by the network simulator. Elites whose performance are closed to the design objectives are selected for subsequent reproduction and mutation. The algorithm iterates until some encoded vectors satisfies the design objectives or the maximum iteration is reached.

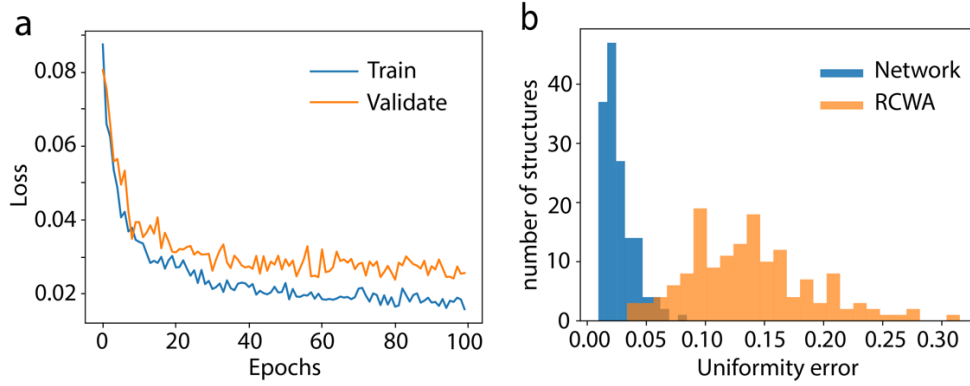

Fig. 4 **Training of the network and statistic of the optimized results** | (a) The variation of training (blue) and validation (orange) loss versus the training epoch. The validation loss reaches 0.03 after 100 epochs of training. (b) Uniformity errors of 150 designed DOE structures with the objective of all diffraction intensities being equal. The blue and orange bars represent the errors calculated with the network simulator and RCWA respectively.

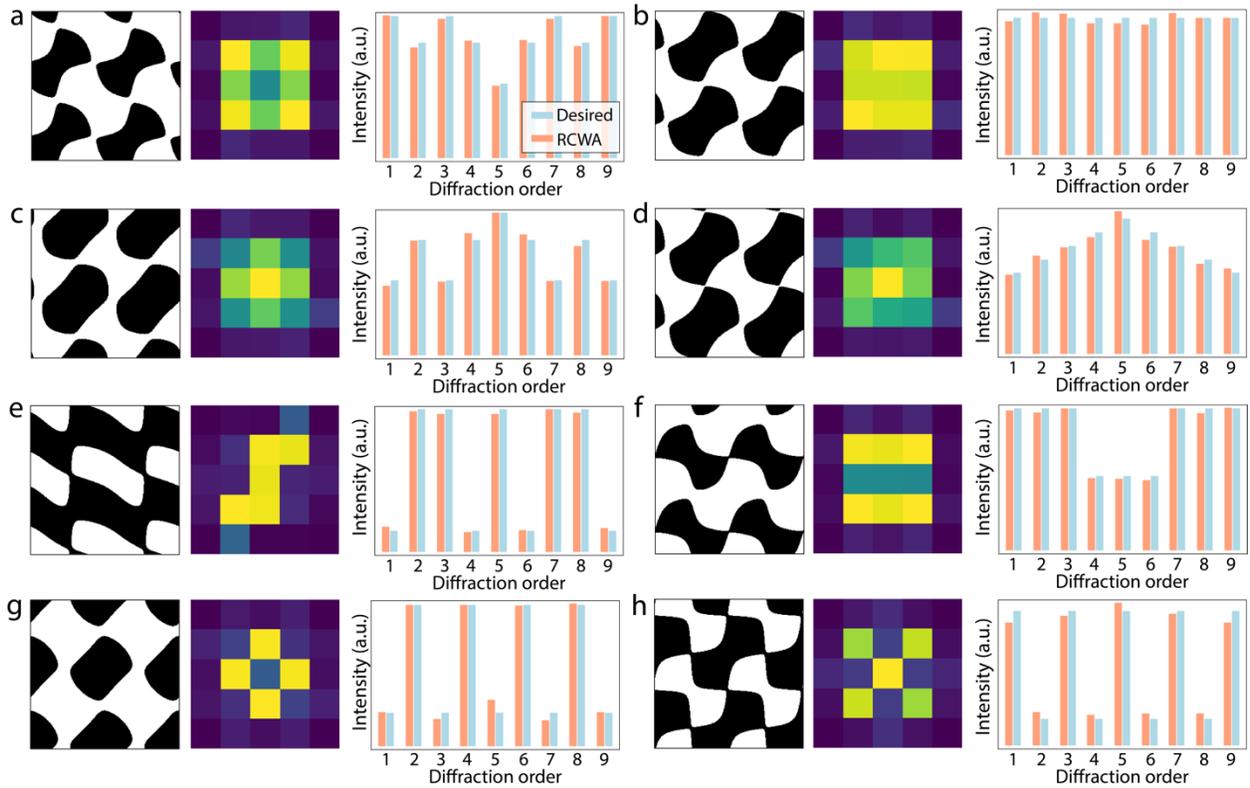

Fig. 4 **Examples of designed DOEs with various diffraction intensity distributions** | In each panel, the leftmost figure presents tiled unit cell of the designed DOE. The middle image represents the simulated efficiencies of all the diffraction orders. Rightmost plot compares the objective intensities (blue) versus the RCWA simulated intensities (orange) of the design. All the designed DOEs are able to diffract light with intensity distributions essentially replicating the design objectives. The uniformity errors of the displayed designs are (a) 0.035, (b) 0.045, (c) 0.073, (d) 0.068, (e) 0.194, (f) 0.036, (g) 0.352, (h) 0.079, respectively. By the definition of the Eq. (9), when the design objectives include diffraction orders with small intensities, tiny disagreement of actual diffraction and objectives induces large uniformity errors. This leads to large $U_{err}$ for the design shown in (e), (g) and (h).